\documentclass[twocolumn,showpacs,preprintnumbers,amsmath,amssymb,nofootinbib]{revtex4}
\usepackage[dvips]{graphicx}
\usepackage{subfigure}
\usepackage{rotating}
\DeclareGraphicsExtensions{.eps}

\def\beq{\begin{equation}}

\def\eeq{\end{equation}}

\def\bea{\begin{eqnarray}}

\def\eea{\end{eqnarray}}
\newcommand{\newc}{\newcommand}

\newc{\ifb}{\textrm{fb}^{-1}}

\newc{\MO}{M_0}

\newc{\Mhalf}{M_{1/2}}

\newc{\AO}{A_0}

\newc{\tanb}{\textrm{tan}\beta}

\newc{\sgnmu}{\textrm{sgn}{\mu}}

\newc{\bbdecay}{0\nu\beta\beta}

\newc{\betadecay}{0\nu\beta\beta}

\newc{\bhalflife}{T^{0\nu\beta\beta}_{1/2}(\Ge)}

\newc{\mee}{m_{\beta\beta}}

\newc{\Ge}{^{76}\textrm{Ge}}

\newc{\bbbar}{B^0_d\textrm{-}\bar{B}^0_d}

\newc{\kkbar}{K^0\textrm{-}\bar{K^0}}

\newc{\solar}{\Delta m^2_{\odot}}

\newc{\atmos}{\Delta m^2_{atm}}

\newc{\mnu}{m_{\nu}}

\newc{\lam}{\lambda}

\newc{\mM}{\mathcal{M}}

\newc{\lag}{\mathcal{L}}

\def\lsim{\raise0.3ex\hbox{$\;<$\kern-0.75em\raise-1.1ex\hbox{$\sim\;$}}}

\def\gsim{\raise0.3ex\hbox{$\;>$\kern-0.75em\raise-1.1ex\hbox{$\sim\;$}}}

\begin{document}
\preprint{DO-TH-10/22}

\title{Lepton number violation in theories with a large number of Standard 
Model copies} 

\author{Sergey~Kovalenko${}^{a}$}
\email[]{sergey.kovalenko@usm.cl}
\author{Heinrich~P\"as${}^{b}$}
\email[]{heinrich.paes@uni-dortmund.de}
\author{Ivan Schmidt${}^{a}$}
\email[]{ivan.schmidt@usm.cl}

\affiliation{${}^{a}$Departamento de F\'{i}sica, 
Universidad T\'{e}cnica Federico Santa Mar\'{i}a\\
and\\
Centro Cient\'{i}fico-T\'ecnologico de  Valpara\'{i}so\\
Casilla 110-V, Valpara\'{i}so,
Valpara\'{i}so, Chile\\
 ${}^{b}$Fakult\"at f\"ur Physik, Technische Universit\"at Dortmund, 
D-44221, Dortmund, Germany}

\date{\today}

\begin{abstract}
We examine lepton number violation (LNV) in theories with a saturated black 
hole bound on a large number of species. Such theories have been 
advocated recently as a possible solution
to the hierarchy problem and an explanation of the smallness of neutrino 
masses.  
The violation of lepton number can be a potential 
phenomenological problem of this 
$N$-copy extension 
of the Standard Model as due to the low quantum gravity scale 
black holes may induce TeV scale
LNV operators generating unacceptably large rates of LNV processes.
We show, however, that this does not happen in this scenario 
due to a specific compensation mechanism between contributions
of different Majorana neutrino states to these processes.
As a result rates of LNV processes
are extremely small and far beyond experimental reach,
at least for the
left-handed neutrino states.
\end{abstract}



\maketitle
\section{Introduction}
\label{sec-1}

Very recently the existence of a large number of copies of 
Standard Model (SM) particles has
been proposed as a possibility to lower the Planck scale 
and solve the electroweak hierarchy problem \cite{Dvali:2007hz,Dvali:2007wp}.

It was shown that in this scenario the fundamental quantum gravity scale 
$\Lambda$ is related to the effective Planck scale $M_{P}$
as
\beq
\Lambda \simeq \frac{M_P}{\sqrt{N}}.
\eeq
This implies $\Lambda \sim {\cal O}(TeV)$ for
$N\simeq 10^{32}$ and thus solves the hierarchy problem  
\cite{Dvali:2007hz,Dvali:2007wp}. 
The above bound is imposed by consistency of large distance black hole dynamics
\cite{Dvali:2007hz,Dvali:2008ec} in the presence of $N$
copies of the SM fields.

Moreover, in \cite{Dvali:2009ne} this scenario has been advocated also
as a mechanism for generating small neutrino masses, providing an attractive 
alternative for seesaw, extra dimensional
and other known mechanisms. 
It is assumed that there exists one SM singlet right-handed neutrino
$\nu_{Rj}$ per SM$_{j}$ copy, so that $j=1,...,N$.  The mechanism relies 
on the fact 
that the right-handed neutrinos, being SM singlets, couple to all the 
SM copies ``democratically''. 
This SM singlet democracy, combined with the requirement of unitarity of 
the theory, 
leads to a $1/\sqrt{N}$ suppression of the Yukawa couplings to 
the left-handed 
neutrinos  $\nu_{Lj}$ and thus a suppression of the 
corresponding Dirac neutrino mass 
terms.
Thus 
the minimalistic approach to the problem of small neutrino masses 
advocated in \cite{Dvali:2009ne} suggests that 
$B-L$ violating Majorana masses 
of $\nu_{Rj}$ are unnecessary in this scenario 
and lepton number could assumed
to be conserved.

The assumption of lepton number conservation is however rather ad hoc,
as there is no fundamental reason to forbid Majorana masses for 
the right-handed neutrinos and lepton number conservation appears  as
an accidental symmetry. Moreover,
quantum gravity breaks global symmetries, and then conserved lepton 
number requires a gauged $B-L$ symmetry 
$U_{1 (B-L)}$. The latter should be spontaneously broken to avoid
the existence of the corresponding massless 
gauge boson, stringently constrained by phenomenology.  
On the other hand, lepton number violation might be 
helpful for successful baryogenesis. 

In the following we analyze the issue of lepton number violation 
within the $N$-copies SM.

\section{Model Framework}
\label{sec-2}
We assume a 
\begin{eqnarray}\label{symmetry}
\prod_{i}\left(SU_{3c}\times SU_{2W}\times U_{y}\right)_{i}\times U_{1 (B-L)}\times Z_{N}
\end{eqnarray}
gauge symmetry of the $N$-copies SM including a common anomaly 
free gauge  factor 
$U_{1 (B-L)}$.  This gauge factor prevents 
the appearance of phenomenologically dangerous 
Lepton Number Violating (LNV) operators induced by TeV black holes.  
An additional permutation symmetry $Z_{N}$ acting in the space of the 
SM$_{i}$ species ($i=1,2,...,N$) is also imposed   \cite{Dvali:2009ne}. 
which, to be unaffected by black holes,  
should be considered as a gauged symmetry in the sense of being a 
discrete subgroup $Z_{N}\subset G$ of some continuous gauge 
group $G$ spontaneously broken down to $Z_{N}$.

The Lagrangian terms relevant for our discussion are the following:
\begin{eqnarray}\label{Lag}
{\cal L}_{\nu H S} = \lambda_{ij} \overline{\nu_{Rj}}\left(L H \right)_{i} + \beta_{ij} \overline{\nu_{Ri}^{c}} \nu_{Rj} S+ \kappa_{i} (H^{\dagger} H)_{i} S^{\dagger} S. 
\end{eqnarray}
The model involves $N$ right-handed SM singlet neutrinos $\nu_{R i}$ 
and one SM singlet complex 
scalar field $S$ having the $B-L$-charge equal to +2. 
Then the trilinear $HHS$ couplings are forbidden in (\ref{Lag}). 
The $U_{1 (B-L)}$ is spontaneously broken by a vacuum expectation value 
$\langle S\rangle$ resulting 
in the appearance of a Majorana mass term from the second term 
in Eq. (\ref{Lag}). 
We assume that the scale of 
$B-L$ breaking lies below the gravity cutoff $\Lambda$. The Dirac mass 
terms considered in Ref. \cite{Dvali:2009ne} 
arise from the first term after the electroweak symmetry breaking.  

We now consider the Yukawa coupling $N\times N$ matrix $\lambda_{ij}$  
of Dirac type, following 
Ref. \cite{Dvali:2009ne}.
As the $\nu_{Ri}$ fields 
are not charged under the SM symmetry, they cannot
be assigned to a single SM-copy, apart from respecting the same transformation
properties under a permutation symmetry acting on the space of species.
This permutation symmetry constrains the Yukawa coupling matrix to the form 
\beq
 \label{lambda}
\lambda_{ij}=\left(
\begin{array}{cccc}
a & b & b & ..\\
b & a & b & ..\\
b & b & a & ..\\
.. & .. & .. & ..\\ 
\end{array}
\right).
\eeq
This matrix, combined in the first term in (\ref{Lag}) with the 
SM Higgs expectation 
value $\langle H_{i}\rangle = \langle H\rangle$, results in the Dirac 
neutrino mass matrix 
\begin{eqnarray}
\label{M-D}
m^{D}_{ij}=\lambda_{ij}\langle H\rangle.
\end{eqnarray}
Here following Ref. \cite{Dvali:2009ne} we assume that the electroweak 
symmetry breaking leaves the permutation symmetry unbroken. This implies 
that the  VEVs of all the Higgs species are equal to the same value 
$\langle H\rangle$.  A key point ensuring in the scenario of 
Ref. \cite{Dvali:2009ne}  the smallness of neutrino mass matrix entries 
is the smallness of the Yukawa coupling matrix (\ref{lambda}), which follows 
from the requirement of unitarity of the theory. This can be shown 
by considering 
right-handed neutrino inclusive production in the scattering of the 
SM particles, as displayed in Fig. 1(a).   
At high energies the rate of this process grows  like
\beq
\label{BD-N}
\Gamma \simeq N b^2 E,
\eeq
as follows from dimensional analysis. Here we assumed $a\sim b$, which 
is suggested by the observation that these two quantities are of the same 
nature and there exists no fundamental reason for them to be 
very different in 
magnitude 
\cite{Dvali:2009ne}.
Unitarity below the gravity cutoff is preserved only for 
\beq\label{b-N}
b\lsim \frac{1}{\sqrt{N}}.
\eeq
Thus the neutrino mass matrix (\ref{M-D}) results  in $N-1$ Dirac 
neutrinos with tiny masses 
$m_D \simeq  \langle H\rangle/\sqrt{N} \lsim {\cal O}(eV)$ \cite{Dvali:2009ne}, 
which fulfill the experimental bounds constraining them to the sub-eV scale. 
One neutrino state in this framework is very heavy, with mass of the order 
$M^{D}\simeq \sqrt{N} \langle H\rangle$, which is comparable with the 
effective
Planck scale.  
\begin{figure}[!t]
    \includegraphics[scale=0.6]{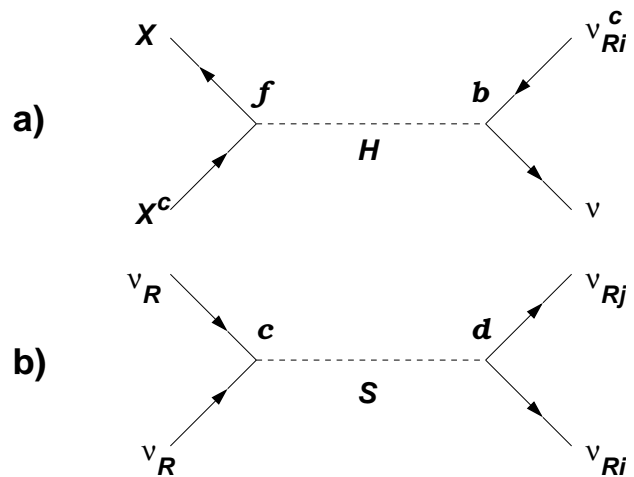}
\caption{Diagrams relevant for the Unitarity constraint on Neutrino Yukawa 
couplings:
(a) via exchange of 
the Higgs doublet $H_{i}$ (Lepton number conserving); (b) via exchange of the
 Higgs 
Singlet 
(Lepton number violating). $X_{i}$ denotes some of the SM fields coupled to 
$H_{i}$ with strength $f$.    
 \label{Large-N}}
\end{figure}

\begin{figure}[!t]
    \includegraphics[scale=0.6]{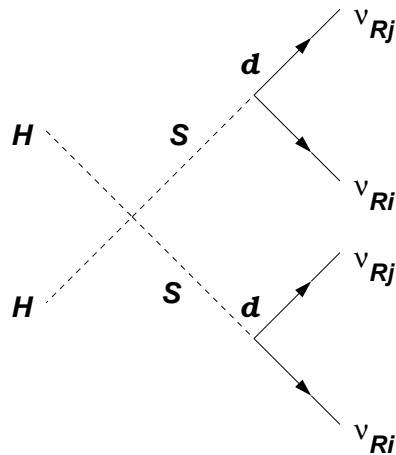}
\caption{The diagram leading to the strongest Unitarity constraint on the 
Majorana neutrino mass term.
 \label{scalar}}
\end{figure}

\begin{figure}[!t]
\smallskip
    \includegraphics[scale=0.6]{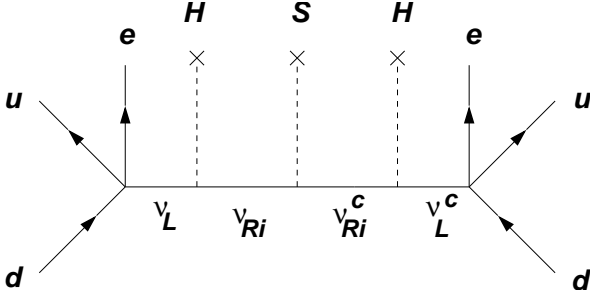}
\caption{Diagram for neutrinoless double beta decay in the presence
of $N$ copies of the SM particle content. 
 \label{0vbb-N}}
\end{figure}

Now let us turn to the second term of Eq. (\ref{Lag}) which upon breaking 
the $B-L$ symmetry leads to the Majorana mass matrix 
of the right-handed neutrinos
\begin{eqnarray}
\label{M-M}
m^{M}_{ij} = \beta_{ij} \langle S\rangle.
\end{eqnarray}
The permutation symmetry  constrains the Majorana type 
Yukawa coupling $N\times N$ matrix 
in the same way as for the case of 
Dirac type Yukawa couplings
to be of the form
\beq
\beta_{ij}=
\left(
\begin{array}{cccc}
c & d & d & ..\\
d & c & d & ..\\
d & d & c & ..\\
.. & .. & .. & ..\\ 
\end{array}
\right).
\eeq
The similarity arguments used above to justify $a\sim b$ in 
Eqs. (\ref{lambda}), 
(\ref{BD-N}) can equally be applied to motivate $c\sim d$. 
Now, when considering the scattering process of right-handed neutrinos,
both final states can be any of the $N$ copies as in Fig. 1(b) and,  
thus, the inclusive rate
grows like  
\beq
\Gamma \simeq N^2 c^2 d^2 E,
\eeq
which preserves unitarity below the gravity cutoff only for 
\beq \label{c-N}
c \sim d \lsim \frac{1}{\sqrt{N}}.
\eeq
An even  more stringent bound results from the diagram of 
Higgs doublet scattering
in Fig.~\ref{scalar},
\beq
\Gamma \simeq N^4 \kappa^{2}d^4 E.
\eeq
The scalar quartic couplings $\kappa_{i}$ should not be very small since 
there is no symmetry protecting its smallness. 
 Thus $\kappa_{i} \sim 1$ can be assumed for rough 
estimations. This implies
\beq\label{cd-lim}
c\sim d \lsim \frac{1}{N}.
\eeq
This limit remains unaffected if additional $S$-branches are inserted 
in the diagram in Fig.~2.
Consequently the
neutrino Majorana mass matrix entries (\ref{M-M}) 
are even more strongly suppressed as the Dirac masses discussed above. 

\section{Neutrino Spectrum and Lepton Number Violation}

We now turn to the neutrino mass spectrum and mixing. 
The mass matrix written in the basis of the 2$N$ fields
${\aleph}_\alpha= \{\nu_{L\beta}, \nu_{R\beta+N}\}$ ($\alpha=1,..., 2N$,
$\beta=1,..., N$)
has the following form,
\begin{eqnarray}\label{MassMatr-1}
{\cal M}^{\nu}=
\left(
\begin{array}{cccc}
0&m^{D}\\
m^{D} &m^{M}\\ 
\end{array}
\right),
\end{eqnarray}
where $m^{D}$ and $m^{M}$ are $N\times  N$ submatrices 
given by Eqs. (\ref{M-D})-(\ref{M-M}).  The set of $2N$ mass eigenstates 
$\nu_{i}=U^\dagger_{i\alpha}{\aleph_\alpha}$ of this symmetric matrix splits into two 
groups of $(N-1)$ degenerate states $\nu^{+}$ and $\nu^{-}$ and 
another two states $N^{\pm}$.  
These groups correspond to two $N-1$ dimensional and two singlet 
representations of  the permutation group $Z_{N}$
\cite{Dvali:2009ne}.
The resulting eigenstates are
\begin{eqnarray}\label{EGS}
&&(N-1)-{\rm plet}:   \  \   \nu^{+}_{k},  \ \ \ k=1,..,N-1    \\
\nonumber
&&\ \ \ \ \ \ \ \ \ \ \ \ \ \ \ \ \ \ \ \ \ \  m_{+}=  \frac{1}{2} 
(\sigma_{0} + \Delta_{0}),\\
 \nonumber
&& \ \ \ \ \ \ \ \ \ \ \ \ \ \  \ \ \              \ \ \ \                    
       U_{1 \nu^{+}_{k}} = \frac{(-1)^{k-1}}{\sqrt{k(k+1)}}
\frac{m_{-}}{\sqrt{ m_{-}^{2} + g_{0}^{2}}},\\
\nonumber
&& (N-1)-{\rm plet}:   \  \    \nu^{-}_{k},     \ \ \ k=1,..,N-1    \\
\nonumber
&&\ \ \ \ \ \ \ \ \ \ \ \ \ \ \ \ \ \ \ \ \ \  m_{- }=  \frac{1}{2} 
(\sigma_{0} - \Delta_{0}),\\
 \nonumber
&& \ \ \ \ \ \ \ \ \ \ \ \ \ \  \ \ \              \ \ \ \                
          U_{1 \nu^{-}_{k}} = \frac{(-1)^{k-1}}{\sqrt{k(k+1)}}
\frac{m_{+}}{\sqrt{m_{+}^{2} + g_{0}^{2}}},\\
\nonumber
 &&\ \ \ \ \ \ \  \  \  {\rm singlet:}       \ \  N^{+} ,     \ \         
M_{+} = \frac{1}{2} (\sigma_{N} + \Delta_{N}), \\ 
 \nonumber
&& \ \ \ \ \ \ \ \ \ \ \ \ \ \  \ \ \              \ \ \ \                    
      U_{1 N^{+}} = \frac{M_{-}}{\sqrt{N (M_{-}^{2} + g_{N}^{2})}},\\
\nonumber
 &&\ \ \ \ \ \ \  \  \ {\rm singlet:}       \ \  N^{-} ,     \ \        
M_{-} = \frac{1}{2}(\sigma_{N} - \Delta_{N}), \\  
 \nonumber
  &&\ \ \ \ \ \ \ \ \ \ \ \ \ \  \ \ \              \ \ \ \              
                U_{1 N^{-}} = \frac{M_{+}}{\sqrt{N (M_{+}^{2} + g_{N}^{2})}}
\end{eqnarray}
with 
\begin{eqnarray}
\nonumber
\sigma_{n} &=& [c + (n-1)d] \langle S\rangle,  \ \ \ 
g_{n} = [a+ (n-1) b]\langle H\rangle, \\
\nonumber
\Delta_{n} &=&  \sqrt{4 g_{n}^{2} + \sigma_{n}^{2}}, \ \ \ \ n=0,N.
\end{eqnarray}
Here we denoted the mixing matrix element 
of each state with the neutrino of the original SM copy by $U_{1a}$. 
All phenomenological 
manifestations of the two sets of  $(N-1)$ degenerate 
states specified above are identical to two effective fields 
$\nu^{+}$ and $\nu^{-}$  with 
masses $m_{+}$ and $m_{-}$, respectively. 
They are defined as
\begin{eqnarray}\label{eff-states}
\nu^{\pm} =  \frac{\sum\limits_{k=1}^{N-1} \nu_k^{\pm} U_{1\nu^{\pm}_k}}{\sum\limits_{k=1}^{N-1}  U^{2}_{1\nu^{\pm}_k}}.
\end{eqnarray}

Thus, the left-handed neutrino 
interaction eigenstate of the original SM copy 
($\alpha=1$)
can be written as 
\begin{eqnarray}\label{Their mixing with the the left handed}
\nu_{L1} = U_{1\nu^{+}}\nu^{+} + U_{1\nu^{-}}\nu^{-} + 
U_{1N^{+}}N^{+} 
+ U_{1N^{-}}N^{-} , 
\end{eqnarray} 
where the effective mixing matrix element of $\nu^{\pm}$ with $\nu_{L1}$ 
is defined as
\begin{eqnarray}\label{U-eff}
U_{1\nu^{\pm}} = \sqrt{\sum_{k=1}^{N-1} U_{1\nu^{\pm}_{k}}^{2}}= 
\frac{m_{\mp}}{\sqrt{m_{\mp}^{2} + g_{0}^{2}}} \sqrt{\frac{N-1}{N}},
\end{eqnarray}
Assuming
$a\sim b$ and $c\sim d$ as discussed above we find
\begin{eqnarray}\label{mass spect}
m_{\pm} &\approx&    \pm m_{\nu}(1\pm \delta_{m})\sim \frac{1}{\sqrt{N}},  \\
\nonumber
M_{\pm} &\approx& \pm M_{N}(1 \pm \delta_{M}) \sim \sqrt{N},
\end{eqnarray}
whith
\begin{eqnarray}\label{def-6}
\nonumber
m_{\nu} &=& (a-b)  \langle H\rangle \sim \frac{1}{\sqrt{N}}, \  
\delta_{m} = \frac{1}{2} \frac{c-d}{a-b} 
\frac{\langle S\rangle}{\langle H\rangle} \sim \frac{1}{\sqrt{N}},\\
M_{N} &=& N b \langle S\rangle \sim \sqrt{N}, \  
\delta_{M} = \frac{1}{2} \frac{d}{b} \frac{\langle S\rangle}{\langle H\rangle}
 \sim \frac{1}{\sqrt{N}}.
\end{eqnarray}
Thus there are two light states $\nu^{\pm}$ and two 
very heavy states $N^{\pm}$ with a mass ratio 
$M_{\pm}/m_{\pm}$
of 
$\sim N$.  For consistency with the neutrino phenomenology one needs one 
neutrino at sub-eV scale, say, 
$m_{\nu} \sim 10^{-2}$eV. Then, 
as seen from Eqs. (\ref{def-6}),
the states $N^{\pm}$ are pushed in mass towards the effective 
Planck scale and, 
therefore, their phenomenological impact is negligible. 
The light Majorana states $\nu^{\pm}$ have a very small mass splitting  
$\delta_{m}\sim 1/\sqrt{N}$ and form a quasi Dirac state 
with mass $m_{\nu}$. Thus these light states 
are expected to induce lepton number 
violating processes at 
rates $\sim 1/N$. However, due to the structure of the mass matrix  
(\ref{MassMatr-1})  with zero submatrix in the upper-left corner, LNV 
processes are even more strongly suppressed. 
The contribution of light $\nu_{k}$ and heavy $N_{k}$ Majorana neutrinos 
with the masses 
$m_{\nu_{k}}$ and $M_{N_{k}}$, respectively,  to  the amplitude
${\cal A}_{LNV}$ of a generic LNV processes 
 can be schematically written as
\begin{eqnarray}\label{LNV-Amplitude expansion}
{\cal A}_{LNV} \sim \sum_{k}\frac{U_{1k}^{2} m_{k}}{p_{0}^{2}+m_{k}^{2}} 
&\approx& \frac{1}{p_{0}^{2}}\langle m_{\nu}\rangle +
\frac{1}{p_{0}^{4}}\langle m_{\nu}^{3}\rangle\\
\nonumber
&+& \left\langle \frac{1}{M_{N}}\right\rangle. 
\end{eqnarray}
Here $m_k$ denotes all mass eigenstates, while
$m_{\nu k}$ and $M_{Nk}$ denote the light and heavy sets  
$m_{\nu k}\ll p_{0}$ and $M_{Nk}\gg p_{0}$, 
respectively, and $p_{0}$ is the 
characteristic momentum of the LNV process under consideration. 
Here we defined
\begin{eqnarray}\label{Def of eff nu masses}
\langle m_{\nu}\rangle &=& \sum_{k}m_{\nu k}U_{1 \nu_{k}}^{2}, \  
\langle m_{\nu}^{3}\rangle = \sum_{k}m_{\nu k}^{3}U_{1\nu_{k}}^{2}, \\
\nonumber 
\left\langle \frac{1}{M_{N}}\right\rangle &=& \sum_{k}
\frac{U_{1N_{k}}^{2}}{M_{N_{k}}}.
\end{eqnarray}
For neutrinoless double beta decay $(0\nu\beta\beta)$, which is the 
most sensitive probe of LNV, the characteristic momentum 
is  $p_{0}\approx$ 105 MeV. For other LNV processes such as 
meson decays and equal sign dileptons in pp-collisions, 
this characteristic momentum $p_{0}$ is even larger. 
 
From the definition (\ref{Def of eff nu masses})  
and Eqs. (\ref{EGS})-(\ref{U-eff}) it follows that 
the leading term of the expansion (\ref{LNV-Amplitude expansion})
vanishes,
$\langle m_{\nu}\rangle = 0$.
This result can be understood by
taking into account  the following two facts. 
First,  the following relation holds, 
\begin{eqnarray}\label{fact-1}
\sum_{k} m_{k} U_{1k}^{2} = {\cal M}^{\nu}_{11} = 0,
\end{eqnarray} 
where now the summation runs over all masses of both heavy and 
light neutrinos. 
Here the mass matrix $ {\cal M}^{\nu}$ with
zero $N\times N$  upper-left corner is given in (\ref{MassMatr-1}). 

Second, as we mentioned before in Eqs. (\ref{EGS}),  the light neutrino 
states belong to two $(N-1)$ dimensional representations of the permutation 
group $Z_{N}$ while the two heavy states are $Z_{N}$-singlets. Thus, due to 
symmetry reasons the cancellation in the sum (\ref{fact-1}) can 
only happen within the same representation, in other words, within 
$(N-1)+(N-1)$ group of light neutrino states and $1+1$
heavy neutrino states.  
This can be directly confirmed by substituting 
Eqs. (\ref{EGS})-(\ref{U-eff}) into Eq. (\ref{Def of eff nu masses}). 

Thus the $N=10^{32}$ SM exhibits the curious property
that
the expression (\ref{fact-1})
vanishes individually also if the sum runs only over the
eigenstates being
much lighter than $p_0$, which corresponds exactly 
to the usually dominating contribution to neutrinoless double
beta decay originating from the first term 
in Eq. (\ref{LNV-Amplitude expansion}).
Consequently
the first non-vanishing contribution 
to the 
LNV amplitude starts 
from the second term in Eq. (\ref{LNV-Amplitude expansion}). This fact can also be illustrated 
diagrammatically as in Fig. \ref{0vbb-N}.  
From Eqs. (\ref{U-eff})-(\ref{def-6}) and (\ref{Def of eff nu masses}) 
it follows that
\begin{eqnarray}\label{m3}
\langle m_{\nu}^{3}\rangle &=& (a-b)^{2}(c-d)\langle H\rangle^{2}\langle S\rangle\sim N^{-2}, \\
\nonumber 
\left\langle \frac{1}{M_{N}}\right\rangle &\approx& \frac{\langle H\rangle}{\langle S\rangle^{2}}\frac{d}{b^{2} N^{2}} \sim  N^{-2}.
\end{eqnarray} 
Therefore, the amplitudes (\ref{LNV-Amplitude expansion}) of LNV processes in the studied framework are extremely small.  

However, this conclusion is based on 
the assumption $a\sim b$ and $c\sim d$ which does not have 
a 
firm physical motivation since the parameters $a, c$ are not directly
subject to the unitarity constraints like in Eq. (\ref{b-N}) 
and (\ref{cd-lim}). We thus 
examine what are the typical rates of LNV processes in the 
scenario in which the latter 
assumption is relaxed. 
Now the parameters $a, c$ should not be expected very small because they 
are not suppressed by
any symmetry. Assuming 
$ \langle S\rangle >  \langle H\rangle$ to avoid phenomenological problems
with a new light gauge boson
in Eqs. (\ref{EGS}) we have
\begin{eqnarray}\label{c-d free}
m_{+} \approx c \langle S\rangle, \ m_{-} \approx \frac{a^{2}}{c} \frac{\langle H\rangle^{2}}{\langle S\rangle},\\
M_{\pm} \approx \frac{1}{2} (c +d N) \langle S\rangle \pm b N \langle H\rangle \sim \sqrt{N}
\end{eqnarray}
Thus, there are still two very heavy states with masses $M_{\pm}\sim \sqrt{N}$, which contribute  to 
the amplitude of LNV processes via the last term in Eq. (\ref{LNV-Amplitude expansion}) and 
this contribution is negligible. The other two Majorana states with masses $m_{\pm}$ are not necessarily 
light, since the suppression due to multiple SM copies is absent now. 
Their masses are related as
\begin{eqnarray}\label{light masses relation}
m_{+} = \left(\frac{\langle S\rangle}{\langle H \rangle}\right)^{2} \left(\frac{c}{a}\right)^{2} m_{-}.
\end{eqnarray}
To ensure the phenomenological consistency of this scenario one has to require that there exists a light neutrino.
Thus, we assume that $m_{-}\sim 10^{-2}$ eV and study the corresponding mass
$m_+$.  The condition $ \langle S\rangle \gg  \langle H\rangle$
is required in order to push the masses of the $B-L$ gauge 
boson and the singlet Higgs $S$ towards sufficiently 
large values to elude current experimental limits.  However in the $N=10^{32}$ SM \cite{Dvali:2009ne} the values of 
$ \langle S\rangle$ larger than the gravity cutoff 
$\Lambda \sim {\cal O}$(TeV)  make little sense. 
Reasonable values are around $ \langle S\rangle \sim 1$~TeV.
In the present scenario there are no symmetry or other reasons supporting any significant difference between the diagonal Yukawa couplings of $H$ and $S$ Higgs fields in Eq. (\ref{Lag}).
Therefore, for estimations it is reasonable to assume $a\sim c$. Then one 
obtains $m_{+} \sim 1$~eV. Thus for this rather typical 
case the  natural values of 
both masses $m_{-}$ and $m_{+}$ 
are much smaller than the typical momentum scales of LNV 
processes $p_{0}\sim 100$ MeV. 
This means that again $\langle m_{\nu}\rangle =0$ and  the leading contribution to the LNV amplitude 
(\ref{LNV-Amplitude expansion}) is 
\begin{eqnarray}\label{m3-2}
{\cal A}_{_{LNV}} p_{0}^{2} &\sim& \frac{1}{p_{0}^{2}}\langle m_{\nu}^{3}\rangle  = m_{-}\left(\frac{m_{+}}{p_{0}}\right)^{2}\\
\nonumber
&\sim& m_{-} \times 10^{-16} \sim 10^{-18}\mbox{eV}. 
\end{eqnarray}
This is far beyond the sensitivity of possible experimental observations. 
The most sensitive $0\nu\beta\beta$ experiments
have reached the limit 
on the double beta decay observable
${\cal A}_{LNV} p_{0}^{2}\sim \langle m_{\nu}\rangle^{exp} \leq 0.38$ eV 
\cite{Strumia:2005tc, Rodin:2007fz}.
This conclusion is valid unless a strong hierarchy  $c\sim 10^{5}a$ between 
the lepton number violating and lepton flavor conserving Yukawa  couplings 
$c$ and $a$, respectively,  is considered. Then $m_{+}\sim 100$ MeV and 
$\langle m_{\nu} \rangle \neq 0$ and
the LNV amplitude (\ref{LNV-Amplitude expansion}) may become large. 
However, as we commented above, this situation is rather unnatural 
since in the present scenario there is no symmetry or other mechanism 
supporting this hierarchy.  

\section{Conclusions}

In this paper we have adressed a problem arising in any scenario
with a low quantum gravity scale: do LNV operators induced by TeV scale
black holes invalidate the model?
For the case of the $N=10^{32}$-copies SM we have shown that this
consequence is avoided due to a non-trivial cancelation mechanism. This
property should be considered as an important benefit of the model.

Nevertheless, the presence of a large number of right-handed Majorana
states may have interesting phenomenological consequences.
For example, a very naive estimate of the right-handed neutrino decay diagrams
on tree and one-loop level, which give rise to the baryon asymmetry in
leptogenesis, scale as $(\sqrt{N})^2$ from the Yukawa coupling with
$N$ copies of $\nu_{Ri}$, contributing potentially to the decay, and the $\nu_{Ri}$
propagator in the loop diagram. So the process may be relevant despite 
the fact that LNV
signals are strongly suppressed for the left-handed neutrino states.
A similar line of reasoning may apply e.g.
to single $\nu_{Ri}$ production
at the LHC. Finally we did not address
the effects of the new neutrino degress of freedom
on big-bang nucleosynthesis \cite{drees}
and the phenomenology of baryon number violation
and proton decay \cite{knochl} in the present scenario.
These and other phenomenological consequences will be discussed
elsewhere.

We conclude that the $N=10^{32}$-copies SM is safe from LNV 
in the SM sector,
and leave the potentially interesting phenomenology of $\nu_{Ri}$
production and decay for further study.

\begin{acknowledgments}
We thank Michele Redi and Gia Dvali for helpful discussions.
This work has been partially supported by the DFG grant PA-803/5-1
and by FONDECYT projects 1100582 and 110287, and 
Centro-Cient\'\i fico-Tecnol\'{o}gico de Valpara\'\i so PBCT ACT-028. 
HP thanks the  Universidad T\'{e}cnica Federico Santa Mar\'{i}a
for hospitality offered while part of this work was carried out. 
\end{acknowledgments}

\end{document}

Naively this result appears to imply that lepton number violation
plays no role in particle physics phenomenology. 
As we will point out in the following, however, this is not the case.

If we estimate e.g. the rate for neutrinoless double beta decay
according to diagram Fig.~\ref{0vbb-N}, the process
is supressed by the small Dirac and Majorana masses but also enhanced
by the large number $N$ of states in the propagator.
Then the effective Majorana mass is given by
\begin{eqnarray}\label{dbd}
m_{\beta\beta}= N m_D \frac{m_M}{p^2} m_D \sim \sqrt{N} 
\frac{m_D^3}{(100~{\rm MeV})^2},
\end{eqnarray}
Here the neutrino momentum being estimated
via the nuclear Fermi momentum $\sim 100$~MeV.
The above equation with  $m_D \simeq {\cal O}(eV)$, 
$m_M \simeq {\cal O}(eV)/\sqrt{N}$
and $N\simeq 10^{32}$ consistent with neutrino oscillation phenomenology predict the effective Majorana 
neutrino mass $m_{\beta\beta}$
in the range presently explored by double beta 
experiments. This non-trivial result is the consequence of a 
stunning coincidence between the mass scale of neutrinos, the
size of the nucleon and the number of states required to lower
the Planck scale down to a TeV.
Another interesting feature of this mechanism is, that the
double beta observable does not coincide with the kinematic neutrino
mass measured in cosmology or Tritium beta decay, which is dominated
by $m_D$. Rather, any neutrino mass bound restricting $m_D$ below
1~eV will supress the neutrinoless double beta decay rate by the third power
of $m_D/(1~{\rm eV})$. This relation may provide a smoking gun for
scenarios with a large number $N\sim10^{32}$ of copies of the SM particle
content.